\let\accentvec\vec
\documentclass[runningheads,a4paper]{llncs}

\let\vec\accentvec
\usepackage{graphicx}
\usepackage{amsmath,psfrag,amssymb}
\usepackage{booktabs} % tables for scientific publishing
\usepackage{longtable}
\usepackage{rotating}
\usepackage{tabularx}       % Fuer Tabellen laenger als eine Seite

\usepackage{url}
\urldef{\mailsa}\path|{boldhaus, klemm}@bioinf.uni-leipzig.de,|
\urldef{\mailsb}\path|{bertschi,jrauh, olbrich}@mis.mpg.de|
\newcommand{\keywords}[1]{\par\addvspace\baselineskip
\noindent\keywordname\enspace\ignorespaces#1}

\newcommand{\inD}[1][\relax]{\def\argone{#1}\def\temprelax{\relax}
  \ifx\argone\temprelax\right.\else\,\middle|#1\right.{}\fi}

\newcommand{\ignore}[1]{}
\newcommand{\ie}{{\it i.e.\ }}
\begin{document}

\mainmatter  % start of an individual contribution

\title{Knockouts, Robustness and Cell Cycles}
\titlerunning{Knockouts, Robustness and Cell Cylces}

\author{Gunnar Boldhaus$^1$ \and Nils Bertschinger$^2$\and Johannes Rauh$^2$\and Eckehard Olbrich$^2$\and Konstantin Klemm$^1$}

\authorrunning{Gunnar Boldhaus et al.}

\institute{$^1$ Department of Bioinformatics, University of Leipzig, Germany\\
$^2$ Max Planck Institute for Mathematics in the Sciences, Leipzig, Germany\\
\mailsa\\
\mailsb\\
\url{http://www.bioinf.uni-leipzig.de}\\
\url{http://www.mis.mpg.de}}

\maketitle

\begin{abstract}
  The response to a knockout of a node is a characteristic feature of
  a networked dynamical system. Knockout resilience in the dynamics of
  the remaining nodes is a sign of robustness. Here we study the
  effect of knockouts for binary state sequences and their
  implementations in terms of Boolean threshold networks. Beside
  random sequences with biologically plausible constraints, we analyze
  the cell cycle sequence of the species {\em Saccharomyces
    cerevisiae} and the Boolean networks implementing it.  Comparing
  with an appropriate null model we do not find evidence that the
  yeast wildtype network is optimized for high knockout
  resilience. Our notion of knockout resilience weakly correlates with
  the size of the basin of attraction, which has also been considered
  a measure of robustness.

\end{abstract}

% ========================================================================
\section{Introduction} 
% ========================================================================

Living systems show an ubiquitous {\em robustness} against mutations,
environmental changes and intrinsic non-determinism
\cite{Elowitz2002,Kitano2004,WagnerBook2005}. In particular, each single
cell must control its growth and eventual division by regulating concentrations
of proteins in a precise temporal pattern. Using a Boolean state dynamics
\cite{Kauffman1969}, this cell cycle network has been argued to be {\em robustly
designed} for budding yeast as a model organism \cite{Li2004}. The robustness
has been pinpointed as reproducibility of the dynamics in the presence of
stochastic perturbations \cite{Li2004,Braunewell2007,Lee2009}.  Resilience
against mutations, \ie changes of the interactions among the proteins, has been
studied as well \cite{Li2004,Lau2007,Boldhaus2009}.

A particular type of mutation, either intrinsic or by intervention, is a
complete knockout of a single protein. Knockouts are a type of component failure
often used in experiments to probe cellular functions. A gene is made
dysfunctional such that it is no longer transcribed and its product is
effectively removed from the cell. Some knockouts can be tolerated or
compensated by the cell, whereas others are lethal. Knockouts are often used to
infer the function of specific proteins. This is only appropriate, if the
knockout is neither lethal nor fully compensated, but disables a specific
function of the cell. Importantly in real experiments, knockout resilient
systems are more difficult to analyze and identify because knockout mutants
do not exhibit measurable difference.  

In this contribution we study the resilience against knockouts in two
scenarios. First, the system under consideration is defined only by a sequence of
activation patterns regardless of the specific mechanism producing them (``black
box''). Here, resilience with respect to knockout of a node means that the
information contained in the activation patterns of all other nodes is still
sufficient to unambiguously produce the original sequence. In the second
scenario, we consider the sequence together with a given implementation by a
Boolean threshold network. Knocking out node $j$ means that we remove its interactions
with the other nodes. The network is resilient against this knockout if all
other nodes still perform the original sequence of activation patterns.

After defining these notions of knockouts and resilience we apply them to the
yeast cell cycle sequence and its network implementations. Significance of the
results is assessed by comparison of random sequences as null models with
various constraints.

%===============================================================================
\section{Defining knockouts and robustness}
%===============================================================================

Molecular processes within cells are frequently modeled by Boolean networks
\cite{Kauffman1969,Davidich2008,Albert2008}. The nodes of the network correspond
to different molecules which can be present in either high or low
concentrations. Interactions between the molecules change these concentrations
which leads to dynamics in discrete time steps of switching events. The dynamics
is thus described as a temporal sequence of activation patterns
$x(0),x(1),\dots$, where each activation pattern $x$ has $n$ binary components
$x_i$, $i = 1,\dots,n$. Low and high concentration of molecule $i$ at time $t$
are denoted by $x_i(t)=0$ and $x_i(t)=1$, respectively. Each node $i$ computes a
Boolean function $f_i$ mapping the present concentration pattern to its
activation at the next time step, \ie $x_i(t+1) = f_i(x_1(t), \ldots, x_n(t))$.

%===============================================================================
\subsection{Robustness of function}
\label{sec:robustness-function}
%===============================================================================

Let us first define what we mean by robustness of a single node $i$ against
knockout of another node $j$. We assume that the input nodes follow a certain
dynamics $x_{1}(t),\dots,x_{n}(t)$, where $t\ge 0$.  The set of possible
activation patterns which the input takes is called the \emph{input support}. 
It can be represented as a set of binary strings of length $n$.

Node $i$ is robust against knockout of node $j$ if $i$ follows the same
sequence even without the presence of node $j$.  This means that the state
information of node $j$ is either irrelevant for $i$ or already contained in the
other available inputs.

\smallskip
\noindent
{\bf Definition:}
\label{def:robustness-definition}
A node $i$ with mapping $f_i: \{0,1\}^n \rightarrow \{0,1\}$ is robust against knockout of node $j$ if $x_i$ is
independent of $x_{j}$ given $x_{1},\ldots,x_{j-1},x_{j+1},\ldots
x_{n}$.  With this we mean that
\begin{equation}
\label{eq:fisf}
  f_i(x_{1},\ldots,x_{j-1},0,x_{j+1},\ldots x_{n})
% \\
  = f_i(x_{1},\ldots,x_{j-1},1,x_{j+1},\ldots x_{n}),
\end{equation}
whenever both $(x_{1},\ldots,x_{j-1},0,x_{j+1},\ldots x_{n})$ and $(x_{1},\ldots,x_{j-1},1,x_{j+1},\ldots x_{n})$ lie in
the input support.
Robustness against simultaneous knockout of multiple nodes is defined in a similar fashion.
\smallskip

It is important to note that this definition depends on the input support.  If the input $x_{1},\ldots, x_{n}$ takes each possible value, then robustness against knockout of node $j$ implies that $f_i$ does not depend on $x_{j}$.
However, if the inputs are highly correlated, then nontrivial robustness may appear: For example, it may be possible that $x_i$ can compensate the knockout of a single input by reconstructing the missing information from another input
with similar dynamics, but the simultaneous knockout of both inputs leads to a failure.

Our notion of knockout robustness is based on ideas of Ay and
Krakauer~\cite{Ay2007}.  Within the framework defined there \cite{Ay2007},
resilience means that the \emph{exclusion dependence} of the system with respect
to certain knockouts vanishes.  Combinatorial conditions characterizing this
situation have been found by Herzog et al.~\cite{Herzog2010}. The theory becomes
much simpler in the present setting restricted to deterministic dynamics.

Our definition of robustness can be applied to each node of the cell cycle
network shown in Figure \ref{fig:cellcyclenetwork}. In our studies of the yeast
cell cycle the input support will be the set of activation patterns which appear in this
cycle. We then study each node mapping and ask, which inputs are ``essential''
for the functioning of this node and which inputs can be compensated.  

In order to study robustness of the system as a whole there are multiple
possibilities: As a
measure of {\em system robustness}
we count the number of nodes which are
robust with respect to all single node knockouts, \ie how many nodes would
remain functional if any one input would be knocked out. Another possibility is
to find the knockouts under which the behavior of the whole system is robust.
In this case we find the single node knockouts which can be compensated by all
(other) nodes of the network. The set of these knockouts is called the
\textit{resilience combination}. The cardinality of the resilience combination
is then used as a measure of {\em knockout resilience}.

%===============================================================================
\subsection{Linear threshold networks and robustness of implementation}
\label{sec:robustness-implementation}
%===============================================================================

The analysis presented up to now just uses properties of the sequence
of node activations and thus cannot distinguish between different protein interaction
networks implementing this sequence. In order to apply our notion of
robustness to example systems and investigate the effects of specific
network structures we focus on linear threshold networks. Such a
network is given as a directed graph among $n$ nodes with weighted
edges. We allow loops, \ie edges starting and ending at the same node.
For simplicity we only allow weights $w_{ij}=1$ and $w_{ij}=-1$ for
each directed edge $i\leftarrow j$. If there is no edge from node $j$
to node $i$ we write $w_{ij}=0$. To each node $i$ we associate a
time-dependent variable $x_{i}(t)$ with dynamics given by
\begin{equation}
  \label{eq:dynamics}
  x_{i}(t+1) =
  \begin{cases}
    1 & \text{ if }\sum_{j} w_{ij}x_{j}(t) > 0,
    \\
    0 & \text{ if }\sum_{j} w_{ij}x_{j}(t) < 0,
    \\
    x_{i}(t) & \text{ else.}
  \end{cases}
\end{equation}

It is possible to refine this model by allowing more values for the
edge weights. Furthermore it is customary to introduce threshold
parameters for every node. Then the dynamics is determined by
comparing $\sum_{j} w_{ij}x_{j}(t)$ to this threshold. In this work we
do not make use of these possibilities and restrict ourselves to the
simple model.

One important feature of the definition of robustness on
page~\pageref{def:robustness-definition} is that it only depends on
the values of the functions $f_i$ on the support.  If we apply this
definition to a Boolean network then, in a sense, the mechanisms at work
are ignored, and only the sequence of activation patterns given by the
dynamics plays a role. Therefore we will call this \emph{robustness of
  function.}

In the context of linear threshold models we can define a related
notion of robustness, which we will call \emph{robustness of
  implementation}: A knockout of a node $j$ is modeled by removing a
node from the network, together with all of the edges involving this
node.  We can then analyze the dynamics of the changed network and
compare it to the dynamics of the unperturbed network.  If the
dynamics of the remaining $n-1$ nodes is not changed, then we say that
the implementation can compensate the knockout of node $j$. This
notion allows to compare different linear threshold networks which
implement the same activation sequence with respect to their knockout
resilience.

%================================================================================
\section{Analysis of the yeast cell cycle}
%================================================================================
\begin{figure}
\centering
\includegraphics[width=0.5\textwidth]{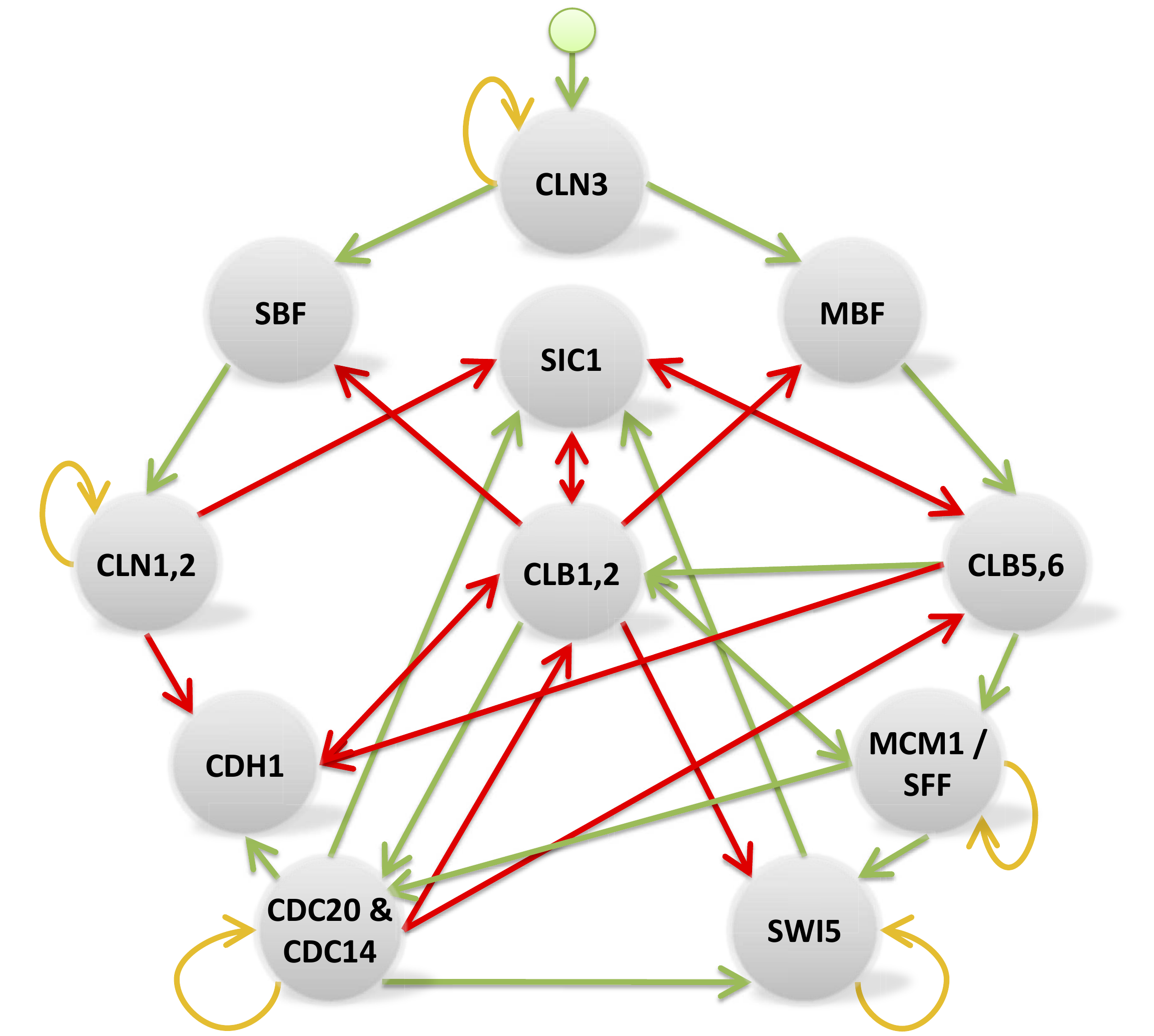}
\caption{\label{fig:cellcyclenetwork}
The wildtype network of the cell cycle of the yeast species
\textit{Saccharomyces cerevisiae} \cite{Li2004}. The edges of the network are
directed and can be activating (green arrow) or inhibiting (red arrow).
All self-couplings (yellow) are inhibiting. The network comprises
$34$ different interactions, $15$ of which are activating and $19$ are
inhibiting.}
\end{figure}

\begin{table}	
  \begin{center}
    \begin{tabular}{ccccccccccccc} \toprule
      \multicolumn{1}{c}{Time} &
      \begin{sideways}CLN3\end{sideways} &
      \begin{sideways}MBF\end{sideways} &
      \begin{sideways}SBF\end{sideways} &
      \begin{sideways}CLN1,2\end{sideways} &
      \begin{sideways}CLB5,6\end{sideways} &
      \begin{sideways}CLB1,2\end{sideways} &
      \begin{sideways}MCM1\end{sideways} &
      \begin{sideways}CDC20\end{sideways} &
      \begin{sideways}SWI5\end{sideways} &
      \begin{sideways}SIC1\end{sideways} &
      \begin{sideways}CDH1\end{sideways} &
      \multicolumn{1}{c}{Phase} \\\midrule
      1  & \textbf{1} & 0 & 0 & 0 & 0 & 0 & 0 & 0 & 0 & \textbf{1} & \textbf{1} & \text{START} \\
      2  & 0 & \textbf{1} & \textbf{1} & 0 & 0 & 0 & 0 & 0 & 0 & \textbf{1} & \textbf{1} & \textrm{$G_1$} \\	
      3  & 0 & \textbf{1} & \textbf{1} & \textbf{1} & 0 & 0 & 0 & 0 & 0 & \textbf{1} & \textbf{1} & \textrm{$G_1$} \\
      4  & 0 & \textbf{1} & \textbf{1} & \textbf{1} & 0 & 0 & 0 & 0 & 0 & 0 & 0 & \textrm{$G_1$} \\
      5  & 0 & \textbf{1} & \textbf{1} & \textbf{1} & \textbf{1} & 0 & 0 & 0 & 0 & 0 & 0 & \textrm{$S$} \\
      6  & 0 & \textbf{1} & \textbf{1} & \textbf{1} & \textbf{1} & \textbf{1} & \textbf{1} & 0 & 0 & 0 & 0 & \textrm{$G_2$} \\
      7  & 0 & 0 & 0 & \textbf{1} & \textbf{1} & \textbf{1} & \textbf{1} & \textbf{1} & 0 & 0 & 0 & \textrm{$M$} \\
      8  & 0 & 0 & 0 & 0 & 0 & \textbf{1} & \textbf{1} & \textbf{1} & \textbf{1} & 0 & 0 & \textrm{$M$} \\
      9  & 0 & 0 & 0 & 0 & 0 & \textbf{1} & \textbf{1} & \textbf{1} & \textbf{1} & \textbf{1} & 0 & \textrm{$M$} \\
      10 & 0 & 0 & 0 & 0 & 0 & 0 & \textbf{1} & \textbf{1} & \textbf{1} & \textbf{1} & 0 & \textrm{$M$} \\
      11 & 0 & 0 & 0 & 0 & 0 & 0 & 0 & \textbf{1} & \textbf{1} & \textbf{1} & \textbf{1} & \textrm{$M$} \\
      12 & 0 & 0 & 0 & 0 & 0 & 0 & 0 & 0 & \textbf{1} & \textbf{1} & \textbf{1} & \textrm{$G_1$} \\
      13 & 0 & 0 & 0 & 0 & 0 & 0 & 0 & 0 & 0 & \textbf{1} & \textbf{1} & \text{Stat.} $G_1$ \\
      \bottomrule
    \end{tabular}
  \end{center}
  \caption{The cell cycle sequence of the yeast species \textit{Saccharomyces cerevisiae} \cite{Li2004}.}
  \label{tab:statetranstion}
\end{table}
For the analysis the model describing the cell cycle of the yeast species
\textit{Saccharomyces cerevisiae}, introduced by Li et al. \cite{Li2004}, is
used. It consists of $11$ nodes (see Figure \ref{fig:cellcyclenetwork}) which
represent four different classes of molecules (cyclins transcription factors and
the inhibitors, degraders, and competitors of the cyclin complexes). Those
molecules are involved in the control of the cell cycle process and act as key
regulators.

%===============================================================================
\subsection{Robustness of function}
%===============================================================================

\begin{figure}
\centering
\includegraphics[width=0.90\textwidth]{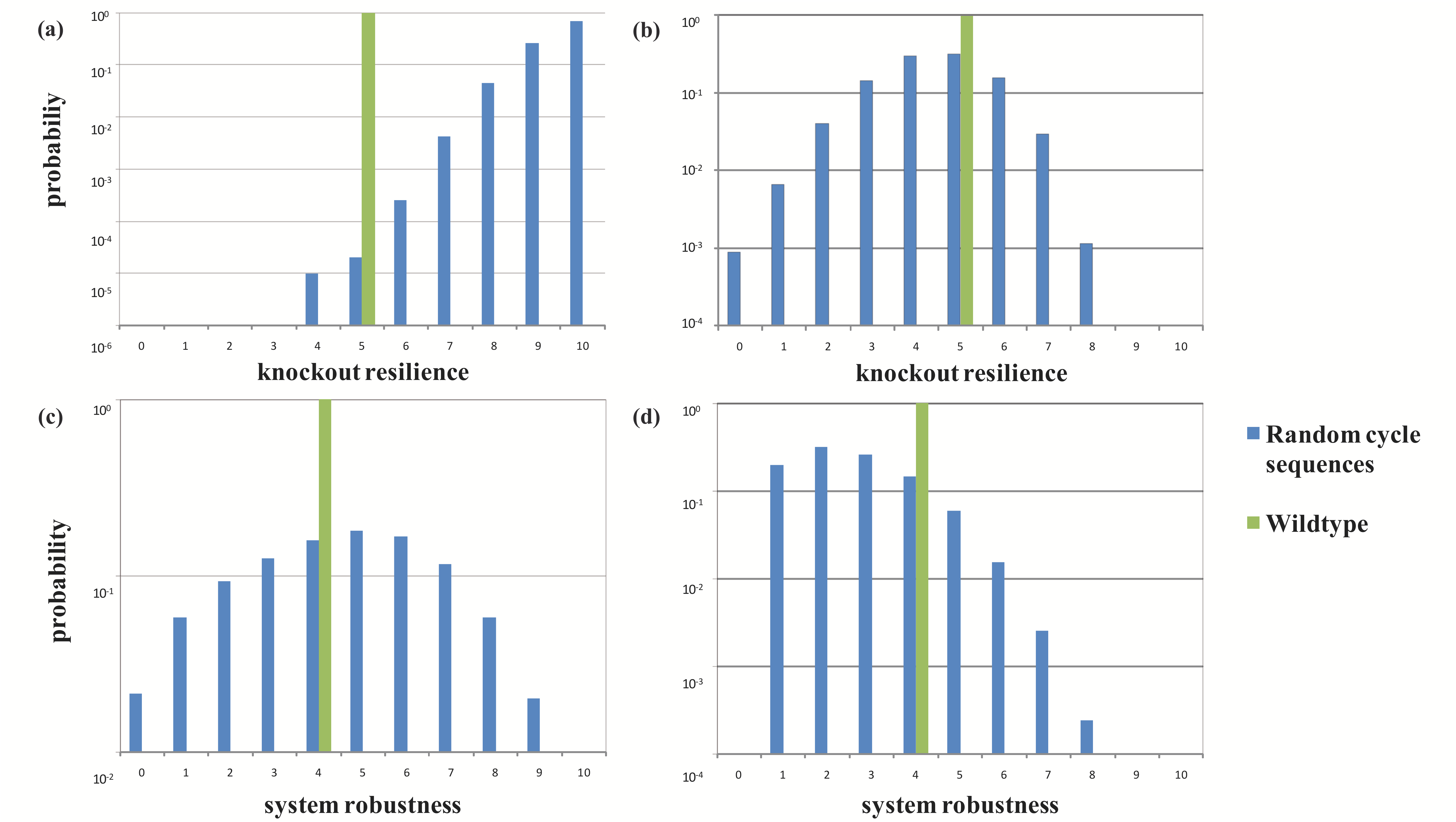}
\caption{$(a-b)$ The number of inputs which can be compensated by all other node are shown for the first null model $(a)$ and the more realistic null model $(b)$. Blue bars represent networks with random cell cycles and the green bar represents the knockout resilience of the wildtype network. $(c-d)$ The number of nodes which are robust with respect to all single node knockouts for the first null model $(c)$ and the more realistic null model $(d)$ are shown. The blue bars represent networks with a random cycle sequence while the green bar shows the system robustness for the wildtype network itself. Note the logarithmic scale of the four diagrams.}
\label{fig:ResultsMPI}
\end{figure}

We now apply the ideas of section~\ref{sec:robustness-function} to the yeast
cell cycle. Starting with the system robustness we find that only the four
nodes \textit{CLN3}, \textit{CLN1,2}, \textit{CLB5,6} and \textit{CDC20} are
robust against all single node knockouts\footnote{For \textit{CLN3} this is
trivial since it corresponds to the constant map $y = 0$.}.
Regarding the knockout resilience the cell cycle still functions correctly if
any one of the five nodes \textit{MBF}, \textit{SBF}, \textit{MCM1/SFF},
\textit{SWI5} or \textit{CDH1} is knocked out.

We can now ask whether these findings are a special property of the cell cycle
or rather expected for a support containing $13$ out of $2^{11} = 2048$ possible
sequences. In order to answer this question we need to define a null model
of sequences which corresponds to a cell cycle. A distinguished feature of the
cell cycle is that it can be triggered by a single signal protein, \ie the first
and last activation pattern only differs by the activity of a single node. As a
first test we generated random activation sequences of $11$ nodes where each
node is active with probability $\frac{1}{2}$. The initial activations were
taken as the last activation pattern with a change in a single node activation.
Cycles containing repeated node activation patterns were discarded and redrawn.

Figure~\ref{fig:ResultsMPI} (a) shows the fraction of random cycles with a
certain knockout resilience. The five nodes of the yeast cell cycle are
untypical compared to this null model, i.e. the yeast can compensate less
knockouts than expected. The system robustness of the random cycles is shown in
Figure~\ref{fig:ResultsMPI} (c). With respect to this measure the four robust
nodes of the yeast cell cycle are rather typical.

Does this mean that the yeast cell cycle is not particularly robust against
knockouts? A closer look at the cell cycle reveals that its activation pattern
is far from random. Another null model which more closely resembles the observed
activation pattern is based on the following properties:
\begin{itemize}
\item \textit{CLN3} acts as an input node. Its activation starts the cycle and
it remains silent throughout the sequence.
\item The other nodes do not switch their activations randomly during the cycle,
but are activated and inactivated exactly once during the cycle. Their activity
(or inactivity) is therefore constrained to a block of successive time steps.
\end{itemize}
Therefore we consider a second null model of random cycles obeying these two
properties. Each node, except node \textit{CLN3} with a fixed sequence, switches
its activation at two randomly drawn time steps within the cycle. As before,
cycles with repeated activation patterns were discarded.

Figure~\ref{fig:ResultsMPI} $(b)$ and $(d)$ show the robustness and knockout
resilience for random cycles from this null model. Now the yeast cell cycle
shows typical values for both considered statistics. This suggests that the
yeast cell might not be optimized for high robustness and knockout resilience,
but just happens to exhibit these properties due to purely statistical reasons.

%===============================================================================
\subsection{Robustness of implementation}
%===============================================================================

\begin{table}
\caption{\label{tab:KOs}
Resilience combinations and their abundance among networks
implementing the cell cycle sequence. {\em All components} refers to
unconstrained implementations while those in the wildtype component
are reachable from the wildtype by one or several mutations.}
		\centering
\begin{tabular}{ccccccccccccccc} \toprule
		\multicolumn{1}{c}{Number} &
		\multicolumn{1}{c}{Wildtype component} &
		\multicolumn{1}{c}{All components} &
		\begin{sideways}CLN3\end{sideways} &
      \begin{sideways}MBF\end{sideways} &
      \begin{sideways}SBF\end{sideways} &
      \begin{sideways}CLN1,2\end{sideways} &
      \begin{sideways}CLB5,6\end{sideways} &
      \begin{sideways}CLB1,2\end{sideways} &
      \begin{sideways}MCM1\end{sideways} &
      \begin{sideways}CDC20\end{sideways} &
      \begin{sideways}SWI5\end{sideways} &
      \begin{sideways}SIC1\end{sideways} &
      \begin{sideways}CDH1\end{sideways} &		
		\multicolumn{1}{c}{\# Knockouts} \\\midrule
1 & 5,13E+25 & 5,07E+34 & $\Box$  & $\Box$ & $\Box$ & $\Box$ & $\Box$ & $\Box$ & $\Box$ & $\Box$ & $\Box$ & $\Box$ & $\Box$ & 0\\
2 & 3,98E+24 & 2,59E+31 & $\Box$ & $\Box$ & $\Box$ & $\Box$ & $\Box$ & $\Box$ & $\Box$ & $\Box$ & $\Box$ & $\Box$ &$\blacksquare$ & 1\\
3 & 6,22E+23 & 1,50E+31 & $\Box$ & $\Box$ &$\blacksquare$ &$\Box$ & $\Box$ & $\Box$ & $\Box$ & $\Box$ & $\Box$ & $\Box$ & $\Box$ & 1\\
4 & 6,22E+23 & 1,50E+31 & $\Box$ &$\blacksquare$ & $\Box$ & $\Box$ & $\Box$ & $\Box$ & $\Box$ & $\Box$ & $\Box$ & $\Box$ & $\Box$ & 1\\
5 &  & 1,77E+31 & $\Box$ & $\Box$ & $\Box$ & $\Box$ & $\Box$ & $\Box$ &$\blacksquare$ & $\Box$ & $\Box$ & $\Box$ & $\Box$ & 1\\
6 &  & 2,24E+29 & $\Box$ & $\Box$ & $\Box$ & $\Box$ & $\Box$ &$\Box$ & $\Box$ & $\Box$ &$\blacksquare$ & $\Box$ &$\blacksquare$ & 2\\
7 &  & 3,15E+32 & $\Box$ & $\Box$ & $\Box$ & $\Box$ & $\Box$ &$\Box$ & $\Box$ & $\Box$ & $\blacksquare$ & $\Box$ & $\Box$ & 1\\
8 &  & 9,67E+28 & $\Box$ & $\Box$ &$\blacksquare$ &$\Box$ & $\Box$ & $\Box$ & $\Box$ & $\Box$ &$\blacksquare$ & $\Box$ & $\Box$ & 2\\
9 &  & 9,67E+28 & $\Box$ &$\blacksquare$ & $\Box$ & $\Box$ & $\Box$ & $\Box$ & $\Box$ & $\Box$ &$\blacksquare$ & $\Box$ & $\Box$ & 2\\
10 &  & 3,52E+27 & $\Box$ &$\blacksquare$ & $\Box$ &$\Box$ & $\Box$ & $\Box$ &$\blacksquare$ & $\Box$ & $\Box$ & $\Box$ & $\Box$ & 2\\
11 & 4,63E+22 & 9,62E+27 & $\Box$ & $\Box$ &$\blacksquare$ &$\Box$ & $\Box$ & $\Box$ & $\Box$ & $\Box$ & $\Box$ & $\Box$ &$\blacksquare$ & 2\\
12 &  & 3,51E+27 & $\Box$ & $\Box$ &$\blacksquare$ & $\Box$ &$\Box$ & $\Box$ &$\blacksquare$ & $\Box$ & $\Box$ & $\Box$ & $\Box$ & 2\\
13 & 4,63E+22 & 9,62E+27 & $\Box$ &$\blacksquare$ & $\Box$ & $\Box$ & $\Box$ & $\Box$ & $\Box$ & $\Box$ & $\Box$ & $\Box$ &$\blacksquare$ & 2\\
14 &  & 8,87E+25 & $\Box$ &$\blacksquare$ & $\Box$ & $\Box$ &$\Box$ & $\Box$ & $\Box$ & $\Box$ &$\blacksquare$ & $\Box$ &$\blacksquare$ & 3\\
15 &  & 8,87E+25 & $\Box$ & $\Box$ &$\blacksquare$ & $\Box$ &$\Box$ & $\Box$ & $\Box$ & $\Box$ &$\blacksquare$ & $\Box$ &$\blacksquare$ & 3\\
16 &  & 1,28E+29 & $\Box$ & $\Box$ & $\Box$ & $\Box$ & $\Box$ & $\Box$  &$\blacksquare$ & $\Box$ &$\blacksquare$ & $\Box$ & $\Box$ & 2\\
17 & 9,12E+20 & 3,22E+24 & $\Box$ &$\blacksquare$ &$\blacksquare$ & $\Box$ &$\Box$ & $\Box$ & $\Box$ & $\Box$ & $\Box$ & $\Box$ &$\blacksquare$ & 3\\
18 &  & 2,44E+22 & $\Box$ &$\blacksquare$ &$\blacksquare$ &$\Box$ & $\Box$ & $\Box$ & $\Box$ & $\Box$ &$\blacksquare$ & $\Box$ &$\blacksquare$ & 4\\
19 &  & 6,25E+23 & $\Box$ &$\blacksquare$ &$\blacksquare$ &$\Box$ & $\Box$ & $\Box$ &$\blacksquare$ & $\Box$ & $\Box$ & $\Box$ & $\Box$ & 3\\
20 &  & 2,79E+25 & $\Box$ &$\blacksquare$ & $\Box$ & $\Box$ & $\Box$ & $\Box$ & $\blacksquare$ & $\Box$ &$\blacksquare$ & $\Box$ & $\Box$ & 3\\
21 & 1,74E+22 & 8,04E+27 & $\Box$ &$\blacksquare$ &$\blacksquare$ & $\Box$ &$\Box$ & $\Box$ & $\Box$ & $\Box$ & $\Box$ & $\Box$ & $\Box$ & 2\\
22 &  & 2,79E+25 & $\Box$ & $\Box$ &$\blacksquare$ &$\Box$ & $\Box$ & $\Box$ &$\blacksquare$ & $\Box$ &$\blacksquare$ & $\Box$ & $\Box$ & 3\\
23 &  & 4,82E+25 & $\Box$ &$\blacksquare$ &$\blacksquare$ &$\Box$ & $\Box$ & $\Box$ & $\Box$ & $\Box$ &$\blacksquare$ & $\Box$ & $\Box$ & 3\\
24 &  & 4,95E+21 & $\Box$ &$\blacksquare$ &$\blacksquare$ &$\Box$ & $\Box$ & $\Box$ &$\blacksquare$ & $\Box$ &$\blacksquare$ & $\Box$ & $\Box$ & 4\\
\textbf{Sum} & \textbf{5,66E+25} & \textbf{5,11E+34} & & & & & & & & & & 
&\\\bottomrule
\end{tabular}
\newline
\end{table}

\begin{table}
\caption{\label{tab:KOTrend}
The average basin sizes for the different knockout resiliences in the wildtype
component and all components are shown.}
	\centering
	\begin{tabular}{ccc} \toprule
	\multicolumn{1}{c}{Knockout Resilience} &
	\multicolumn{1}{c}{Wildtype Component} &
	\multicolumn{1}{c}{All Components}\\\midrule
0 & 1663,44 & 1447,42\\
1 & 1629,86 & 1485,06\\
2 & 1675,99 & 1513,21\\
3 & 1779,22 & 1588,93\\
4 &  & 1594,82\\\bottomrule
	\end{tabular}
\newline
\end{table}

\begin{figure}
\centering
\includegraphics[width=0.99\textwidth]{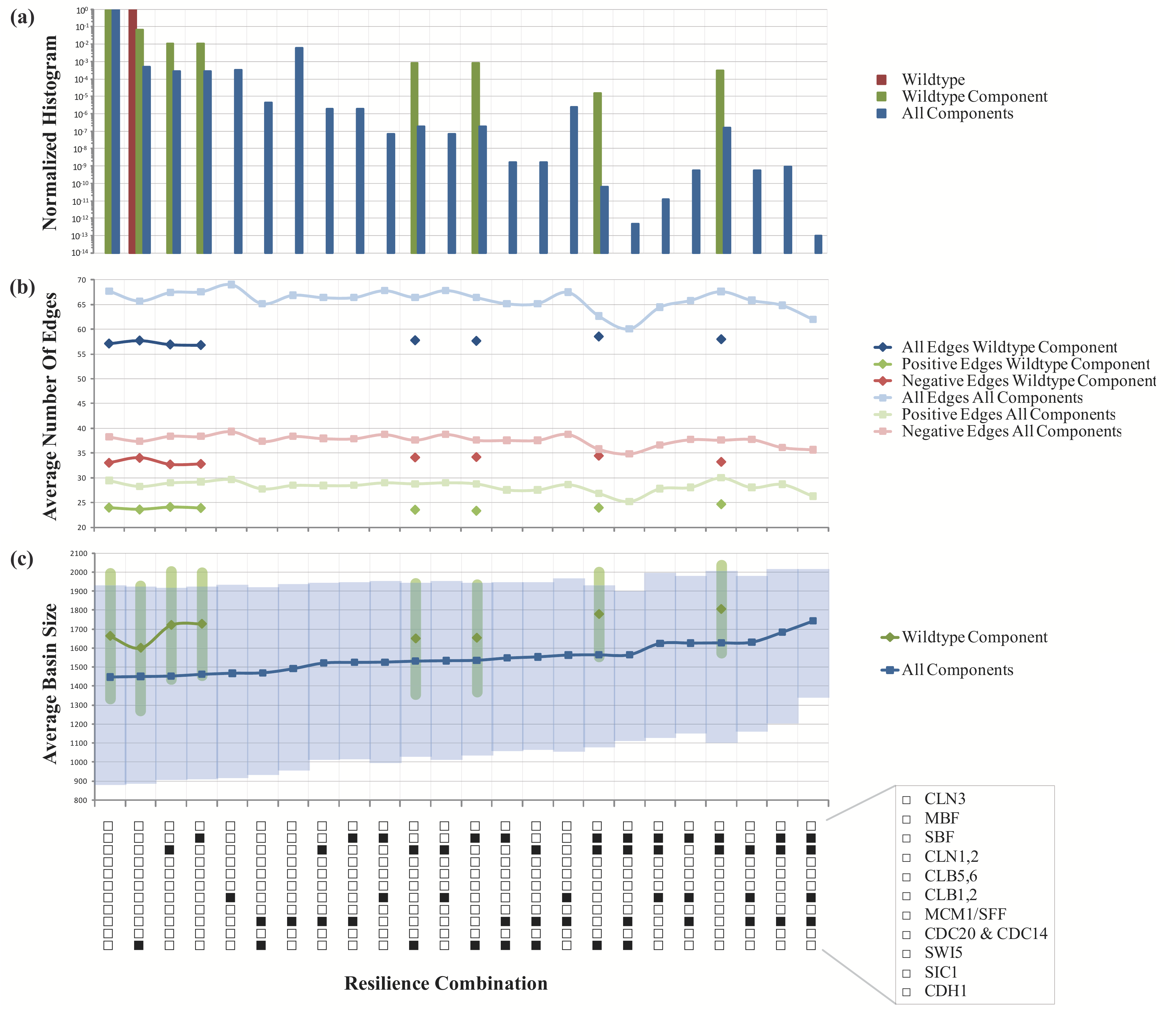}
\caption{\label{fig:OverviewKO}
The three diagrams show different aspects of the knockout experiment. One column
represents one specific resilience combination shown on the bottom. 
$(a)$ The distribution of the specific knockout schemes is shown. The green bars
represent the wildtype component. The blue bars represent all components and the
single red bar represents the resilience combination of the wildtype itself.
Altogether $24$ different resilience combinations were found, $8$ of which are
only present in the wildtype component. $(b)$ The distribution of positive,
negative and all edges for each of the corresponding resilience combinations are
shown. For each resilience combination $10^5$ network were sampled. The darker
lines with the squares represent networks from all components, while the
brighter lines with diamonds represent networks only reachable by mutations from
the wildtype. $(c)$ For each knockout scheme $10^5$ different networks were
sampled. The averages of the basin sizes corresponding to the specific
resilience combination are shown. The blue line with squares represents the
networks from all components, while the green line with a diamond represents
networks only from the wildtype component. Additionally, the area between the
$10\%$ and $90\%$ quantiles is inked. The
resilience combinations are ordered such that the average basin size increases.}
\end{figure}

To further understand to what extent knockout resilience played a role for the
evolutionary design of the cell cycle, we investigated its implementation as
linear threshold networks. For the knockout experiment we looked at all networks
which are capable of reproducing the cell cycle sequence (see Table
\ref{tab:statetranstion}). The set of all such networks can be considered as the
set of vertices of the \emph{neutral graph}. Two networks are then connected by
an undirected edge if they differ by a single mutation. In our case a mutation
corresponds to a addition or deletion of an edge or the change of a weight in
the interaction network \cite{Boldhaus2009,Schuster:94a}. 

As shown in \cite{Boldhaus2009} the neutral graph decomposes into many different
connected components. This is why we compare the wildtype network with the
wildtype component as well as the whole neutral graph. For each network we
checked if each possible single node knockout is changing the cell cycle
sequence (see Table \ref{tab:statetranstion}) or if it has no influence on it. 
We found $24$ different occurring resilience combinations. Out of these only
eight were found in the component where the wildtype is situated (see Table
\ref{tab:KOs}). 

The majority of networks, \ie $\approx 99\%$ of all the functional networks of
the neutral graph and $\approx 90\%$ of the networks in the wildtype component,
cannot cope with any single node knockout. However, a maximum of four
independent single node knockouts were found. The wildtype itself is found to be
only capable of coping with the single node knockout of the node \textit{CDH1},
see Figure \ref{fig:OverviewKO} $(a)$ and networks reachable by a mutational
path from the wildtype can manage at maximum three independent knockouts.

One might speculate that high knockout resilience requires redundant wiring of
the network which would be observable as an increased edge density. In Figure
\ref{fig:OverviewKO} $(b)$ we look at the distribution of the number of
positive, negative and all edges for the different resilience combinations.
There is no clear correlation between the average number of edges and the
knockout resilience.

As suggested by Li et al. \cite{Li2004}, the basin of attraction of the G1 fixed
point (stationary state) is a measure of robustness. The basin consists of all
activation patterns from which the G1 state is eventually reached by following
the dynamics (\ref{eq:dynamics}). The average basin size of networks with a
given resilience combination is shown in Figure \ref{fig:OverviewKO} $(c)$. 
In Table \ref{tab:KOTrend} the average basin size
for the different knockout resiliences is shown. With an
increasing capability to cope with more single node knockouts the average basin
size increases. Additionally, average basins sizes in the wildtype component are
larger then their corresponding basin sizes in all components. However, all network 
implementations in the wildtype component have a network resilience of at most three.

% ========================================================================
\section{Discussion}
% ========================================================================

We have studied dynamics with Boolean state vectors (activation patterns) under
knockout of single components, being the nodes of an underlying network.  The
yeast wildtype network is not optimized for knockout resilience, given the
sequence of activation patterns. There are networks with significantly larger
knockout resilience implementing the same sequence. 

Finally, we stress that our definition of resilience checks whether a node is
dispensable for integrating and transmitting information only in the context of
the regulatory network we consider. This regulatory network is far from being a
complete description of cell function. In reality, therefore, the considered
node may be involved in other functions and indispensable for survival thereby
further reducing the knockout resilience.

%===============================================================================
\subsubsection*{Acknowledgments.}
%===============================================================================

This work was supported by the Volkswagenstiftung. We are grateful to Nihat Ay
for useful comments.

\bibliographystyle{splncs}
\bibliography{literature}

\begin{thebibliography}{10}

\bibitem{Elowitz2002}
Elowitz, M.B., Levine, A.J., Siggia, E.D., Swain, P.S.:
\newblock {Stochastic gene expression in a single cell}.
\newblock Science (New York, N.Y.) \textbf{297}(5584) (2002)  1183--6

\bibitem{Kitano2004}
Kitano, H.:
\newblock {Biological robustness}.
\newblock Nature reviews. Genetics \textbf{5}(11) (2004)  826--37

\bibitem{WagnerBook2005}
Wagner, A.:
\newblock Robustness and Evolvability in Living Systems.
\newblock Princeton University Press (2005)

\bibitem{Kauffman1969}
Kauffman, S.:
\newblock Metabolic stability and epigenesis in randomly constructed genetic
  nets.
\newblock Journal of Theoretical Biology \textbf{22}(3) (Mar 1969)  437--467

\bibitem{Li2004}
Li, F., Long, T., Lu, Y., Ouyang, Q., Tang, C.:
\newblock {The yeast cell-cycle network is robustly designed}.
\newblock Proceedings of the National Academy of Sciences of the United States
  of America \textbf{101}(14) (2004)  4781--6

\bibitem{Braunewell2007}
Braunewell, S., Bornholdt, S.:
\newblock Superstability of the yeast cell-cycle dynamics: Ensuring causality
  in the presence of biochemical stochasticity.
\newblock Journal of Theoretical Biology \textbf{245}(4) (2007)  638 -- 643

\bibitem{Lee2009}
Lee, W.B., Huang, J.Y.:
\newblock {Robustness and topology of the yeast cell cycle Boolean network}.
\newblock FEBS letters \textbf{583}(5) (2009)  927--32

\bibitem{Lau2007}
Lau, K.Y., Ganguli, S., Tang, C.:
\newblock {Function constrains network architecture and dynamics: A case study
  on the yeast cell cycle Boolean network}.
\newblock Physical Review E \textbf{75}(5) (Mai 2007)  1--9

\bibitem{Boldhaus2009}
Boldhaus, G., Klemm, K.:
\newblock {Regulatory networks and connected components of the neutral space}.
\newblock (2009)

\bibitem{Davidich2008}
Davidich, M.I., Bornholdt, S.:
\newblock Boolean network model predicts cell cycle sequence of fission yeast.
\newblock PLoS ONE \textbf{3}(2) (2008)  e1672

\bibitem{Albert2008}
Albert, I., Thakar, J., Li, S., Zhang, R., Albert, R.:
\newblock {Boolean network simulations for life scientists}.
\newblock Source code for biology and medicine \textbf{3} (Januar 2008) ~16

\bibitem{Ay2007}
Ay, N., Krakauer, D.C.:
\newblock Geometric robustness theory and biological networks.
\newblock Theory in Biosciences \textbf{125}(2) (2007)  93 -- 121

\bibitem{Herzog2010}
Herzog, J., Hibi, T., Hreinsdóttir, F., Kahle, T., Rauh, J.:
\newblock Binomial edge ideals and conditional independence statements.
\newblock Advances in Applied Mathematics (2010) ~-- In Press, Corrected Proof.

\bibitem{Schuster:94a}
Schuster, P., Fontana, W., Stadler, P.F., Hofacker, I.L.:
\newblock From sequences to shapes and back: A case study in {RNA} secondary
  structures.
\newblock Proc.\ Roy.\ Soc.\ Lond.\ B \textbf{255} (1994)  279--284

\end{thebibliography}

\end{document}